\documentclass[conference]{IEEEtran}
\IEEEoverridecommandlockouts
\usepackage{cite}
\usepackage{amsmath,amssymb,amsfonts}
\usepackage{algorithmic}
\usepackage{graphicx}
\usepackage{textcomp}
\usepackage{xcolor}
\usepackage{tikz}
\usepackage{comment}
\usepackage{stfloats}  
\def\BibTeX{{\rm B\kern-.05em{\sc i\kern-.025em b}\kern-.08em
    T\kern-.1667em\lower.7ex\hbox{E}\kern-.125emX}}
\usepackage{enumitem,amssymb}
\usepackage{placeins}

\usepackage{threeparttable}
\usepackage{booktabs}
\usepackage{array}

\usepackage[table]{xcolor}

 \usepackage{tikz}

\newlist{todolist}{itemize}{2}
\setlist[todolist]{label=$\square$}

\newcommand{\review}[1]{{\textcolor{black}{#1}}}
\begin{document}
\title{Wearable and Ultra-Low-Power Fusion of EMG and A-Mode US for Hand-Wrist Kinematic Tracking\\
\thanks{The authors acknowledge support from the ETH Research Grant \mbox{ETH-C-01-21-2} (Project ListenToLight).}
}
\author{
\IEEEauthorblockN{
    Giusy Spacone\IEEEauthorrefmark{1}, 
    Sebastian Frey\IEEEauthorrefmark{1}, 
    Mattia Orlandi\IEEEauthorrefmark{2},
    Pierangelo Maria Rapa\IEEEauthorrefmark{2}, 
    Victor Kartsch\IEEEauthorrefmark{1}, \\
    Simone Benatti\IEEEauthorrefmark{3}, 
    Luca Benini\IEEEauthorrefmark{1}\IEEEauthorrefmark{2}, 
    Andrea Cossettini\IEEEauthorrefmark{1}
}
\IEEEauthorblockA{\IEEEauthorrefmark{1}
Integrated Systems Laboratory, ETH Zurich, Z{\"u}rich, Switzerland, \IEEEauthorrefmark{2}DEI, University of Bologna, Bologna, Italy,\\\IEEEauthorrefmark{3} DIEF, University of Modena and Reggio Emilia, Reggio Emilia, Italy}
}

\makeatletter
\def\ps@IEEEtitlepagestyle{%
  \def\@oddhead{}
  \def\@evenhead{}%
  \def\@oddfoot{%
    \vbox to0pt{\vss
      \hbox to\textwidth{%
        \parbox[t]{\textwidth}{\centering\scriptsize
          \copyright 2025 IEEE.  Personal use of this material is permitted.  Permission from IEEE must be obtained for all other uses, in any current or future media, including reprinting/republishing this material for advertising or promotional purposes, creating new collective works, for resale or redistribution to servers or lists, or reuse of any copyrighted component of this work in other works.
        }%
      }%
    }%
  }%
  \def\@evenfoot{}%
}
\makeatother

\maketitle

\vspace{-1cm}  
\begin{abstract}
Hand gesture recognition based on biosignals has shown strong potential for developing intuitive human–machine interaction strategies that closely mimic natural human behavior. In particular, sensor fusion approaches have gained attention for combining complementary information and overcoming the limitations of individual sensing modalities, thereby enabling more robust and reliable systems. Among them, the fusion of surface electromyography (EMG) and A-mode ultrasound (US) \review{is very promising.} However, prior solutions rely on power-hungry platforms unsuitable for multi-day use and are limited to discrete gesture classification.
In this work, we present an ultra-low-power (sub-50 mW) system for concurrent acquisition of 8-channel EMG and 4-channel A-mode US signals, integrating two state-of-the-art platforms into fully wearable, dry-contact armbands. We propose a framework for continuous tracking of 23 degrees of freedom (DoFs), 20 for the hand and 3 for the wrist, using a kinematic glove for ground-truth labeling. Our method employs lightweight encoder–decoder architectures with multi-task learning to simultaneously estimate hand and wrist joint angles. Experimental results under realistic sensor repositioning conditions demonstrate  \review{that EMG--US fusion achieves a root mean squared error of $10.6^\circ\pm2.0^\circ $, compared to $12.0^\circ\pm1^\circ$ for EMG and $13.1^\circ\pm2.6^\circ$ for US, and a  R$^2$ score of $0.61\pm0.1$, with $0.54\pm0.03 $ for EMG and $0.38\pm0.20$ for US}. 


\end{abstract}
\vspace{-0.2cm}
\section{Introduction}
\vspace{-0.1cm}

Hand Gesture Recognition (HGR) is a widely explored topic for developing novel Human-Machine Interaction (HMI) strategies~\cite{guo_human-machine_2021}, with applications spanning Extended Reality (XR), teleoperation, alternative control interfaces, and assistive technologies \cite{tchantchane_review_2023}. Electromyography (EMG), the predominant biosignal modality for HGR \cite{shin_hgrewview_2024}, faces limitations such as poor resolution of individual muscle fiber dynamics \cite{boyer_emgnoise_2023}, susceptiblity to noise, and inter/intra-subject variability \cite{shin_hgrewview_2024}.

Ultrasound (US) has emerged as a complementary modality, thanks to its ability to capture muscle dynamics at varying depths and to precisely track changes in muscle thickness during movement \cite{yang_simultaneous_2022}. Although early efforts focused on benchtop B-mode US, recent work \cite{yang_simultaneous_2022, grandi_sgambato_high_2023, frey_biogap_2023} has focused on wearable \mbox{A-mode} platforms for non-invasive, continuous monitoring. 

Combining EMG and US can leverage their complementary electromechanical characteristics, mitigating individual limitations, and improving robustness. Notable results include EMG-US HGR across 20 gestures with up to $94.14\%$ accuracy \cite{xia_toward_2019,wei_multimodal_2023}, simultaneous HGR and force measurement \cite{yang_comparative_2020}, and use in both healthy participants and transradial amputees \cite{yin_wearable_2025}.

However, most of the previous works have focused on discrete classification. Yet, for intuitive HMI, continuous estimation of multiple hand/wrist degrees of freedom (DoFs) is needed. Though EMG~\cite{krasoulis_effect_2019, zanghieri_semg-based_2021, liu_neuropose_2021, salter_emg2pose_2024, tacca_wearable_2024, lin_continuous_2025} and US~\cite{yang_wearable_2021, yang_simultaneous_2022, grandi_sgambato_high_2023, spacone_tracking_2024, sgambato_multidofprosthetic_2024} have independently been explored for regression, to the best of our knowledge,  no studies have combined them for wearable HGR regression.
Two key challenges need to be addressed to enable multimodal HGR regression in wearable devices:(\textit{i}) the development of multimodal (EMG+US) and low-power wearable platforms, and  (\textit{ii}) efficient sensor fusion algorithms.
Current systems, such as an 8-ch. EMG + 8-ch. US with  Ethernet links \cite{xia_toward_2019}, or a 4-ch. EMG + 4-ch. US with a Cortex-A7 MCU \cite{yin_wearable_2025} exceed the energy envelope suitable for continuous multi-day operations or exhibit limited onboard processing, hindering real-time, low-latency applications.

Concerning the sensor fusion algorithms, early works ~\cite{xia_toward_2019, yang_comparative_2020, yin_wearable_2025} used hand-crafted features with dimensionality reduction, while recent deep-learning approaches \cite{zeng_feature_2020, wei_multimodal_2023, pan_msmfnet_2024, zhang_dual-modal_2025} show better performance, also enabling cross-modal interaction and hybrid feature generation. Still, there remains a need for accurate and computationally efficient fusion algorithms tailored for edge computing, supporting low-latency inference, real-time processing, and operating within the memory and power constraints of wearable platforms.
To bridge the gaps identified at both the platform and algorithmic levels, we put forward the following contributions:
\begin{itemize}
    \item Ultra-low-power (sub-$50\,\text{mW}$) system for concurrent \mbox{8-ch.} EMG and 4-ch. A-mode US data acquisition, integrating two SoA platforms \cite{frey_biogap_2023, frey_wulpus_2022} into fully wearable, dry-contact armbands.
    \item Development of a new framework for concurrent hand (20 DoFs) and wrist (3 DoFs) kinematic tracking based on EMG and A-mode US, using a kinematic glove for ground-truth labeling.
    \item Proposal of lightweight encoder-decoder network architectures leveraging multi-task learning for simultaneous estimation of hand and wrist joint angles.
    \item Demonstration of root mean squared error with sensor fusion as low as $10.6^\circ\pm2.0^\circ $, compared to $12.0^\circ\pm1^\circ$ for EMG and $13.1^\circ\pm2.6^\circ$ for US, and a R$^2$ score of $0.61\pm0.1$, with $0.54\pm0.03 $ for EMG and $0.38\pm0.2$ for US under the realistic sensor repositioning scenario. 
\end{itemize}

\section{Materials and Methods}
\vspace{-0.1cm}
\subsection{Electronic platform for EMG-US data acquisition}
\vspace{-0.1cm}

\begin{figure*}
    \centering
    \includegraphics[width=0.98\textwidth]{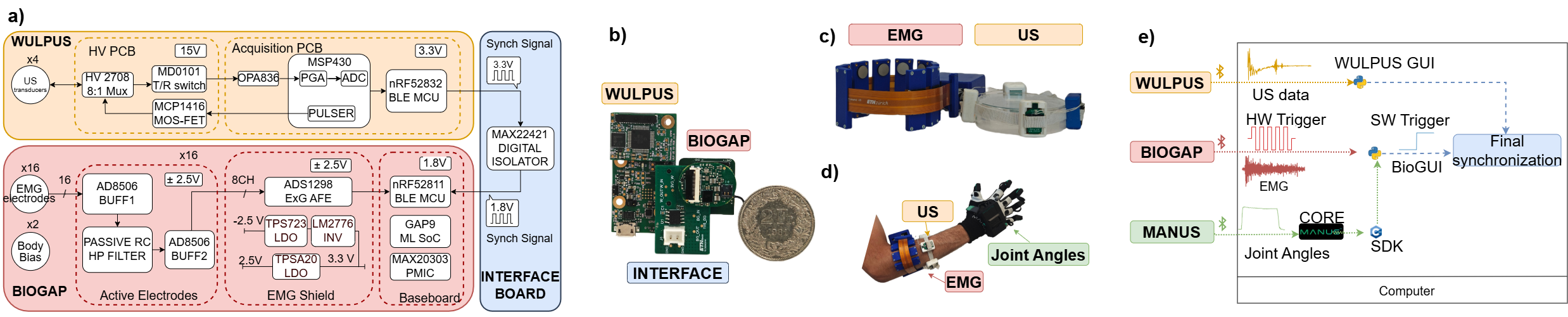}
    \vspace{-0.3cm}
    \caption{a) Block diagram of EMG-US integrated Platform; b) Integrated Hardware c) EMG and US dry armbands d) Data collection framework e) Overview of the EMG-US-MANUS synchronization procedure. US signals are acquired by the WULPUS GUI, whereas glove data are streamed to the MANUS Core and then forwarded to the BioGUI, which also acquires EMG and the hardware synchronization signal from BioGAP. BioGAP and MANUS data are synchronized using the software-based trigger generated by the BioGUI, and then they are synchronized with WULPUS data using the hardware synchronization signal. }
    \vspace{-0.6cm}
    \label{fig:integrated_platform}
\end{figure*}

Fig.~\ref{fig:integrated_platform} (a-b) shows our proposed system for concurrent EMG and US data acquisition, based on SoA acquisition platforms for EMG (BioGAP \cite{frey_biogap_2023}) and A-mode US (WULPUS \cite{frey_wulpus_2022}), extending the proof-of-concept interfacing work of \cite{frey_ius2023}.

The WULPUS subsystem is extensively described in \cite{frey_biogap_2023}. The BioGAP subsystem features a baseboard for digital signal processing \cite{frey_biogap_2023} and a novel 8-channel (differential) expansion board for EMG data collection with an ADS1298 analog front-end \cite{ads_1298} and a balanced power supply ($\pm2.5~\text{V}$). 
The 8 EMG channels are configured in fully differential mode to maximize noise rejection and minimize interference due to the US switching activity. 
To improve signal-to-noise ratio (SNR), we designed custom active EMG electrodes (operating at $\pm2.5~\text{V}$) with a signal buffering stage (AD8605 \cite{AD8605_amplifier}, Analog Devices), followed by a passive high-pass filtering stage ($15~\text{Hz}$ cut-off), together with a second buffering stage.

An interface PCB connects the two subsystems via a two-channel digital isolator (MAX22421 \cite{MAX22420_digital_isolator}) to minimize interference and crosstalk. The US microcontroller unit (MCU) acts as a synchronization master and generates a 3.3 V trigger signal every 50 US excitation pulses. This trigger is received through the isolator by a GPIO pin on BioGAP's nRF52 MCU (operating at 1.8 V) and is appended to the corresponding EMG data packets. 
Data from the two platforms are then streamed independently to a laptop via BLE and are time-aligned in software (Python).

\vspace{-0.1cm}
\subsection{EMG and US armbands}
\vspace{-0.1cm}

The US-EMG electronic acquisition platform is connected to two separate armbands for EMG and US data acquisition.

\textbf{EMG:} we designed an armband with 16 SoftPulse dry EMG electrodes \cite{datwyler_softpulse} (8 differential channels) and an additional electrode pair for bias. The electrodes are connected to the active electrode PCB using a metal snap connected and housed inside 3D plastic enclosures, offering protection to electrical components.  A flex PCB connects the electrodes to the acquisition electronics. The armband comes in three sizes, to be used by individuals with different forearm circumferences. 

\textbf{US:} we use the same armband as in \cite{spacone_tracking_2024}, with four 32-element linear array transducers (Vermon) with a central frequency of $2.25~\text{MHz}$, equally spaced around the forearm. For each transducer, the 4 centermost channels are shorted together and used as an equivalent single-channel element. Hydrogel pads \cite{hydrogel_pads} are used for acoustic coupling, as they do not require continuous reapplication and remain adherent to the skin during use (minimizing transducer displacement).

The EMG armband is placed proximal to the elbow, with the US armband placed more distally (first half of the forearm). Figure \ref{fig:integrated_platform}(d) shows the data collection set-up, where a kinematic glove is used as ground truth (see next section).
\vspace{-0.2cm}
\subsection{Data collection framework: EMG, US, and kinematic glove}
\vspace{-0.1cm}
\textbf{Ground truth}. The Manus Quantum Metaglove \cite{manus_quantum_glove} (MANUS, Netherlands) is used as ground truth for wrist/hand kinematics. It features a 9-axis IMU positioned on the back of the hand and a pair of EM field receivers-transmitters positioned in finger sleeves. The source generates an electromagnetic field, and the detectors capture its variations induced by finger movements \cite{saggio_quasi-static_2025}. Raw data acquired by the glove is transmitted via BLE to the MANUS Core software running on a local laptop. A proprietary algorithm calculates 20 finger joint angles: for each finger, the spread of the carpometacarpal joint (CMC), the flexion of the CMC, the flexion of the proximal interphalangeal joint (PIP) and the flexion of the distal interphalangeal joint (DIP). To represent the 3-axis wrist rotation, the algorithm computes quaternions, acceleration, and absolute heading from the 9-axis IMU data. In addition, the Core provides timestamps. The MANUS SDK \cite{manus_quantum_glove}, developed in C++, provides an interface to access MANUS Core data. 

\textbf{Data synchronization and protocol management}. We expanded the Python-based BioGUI~\cite{biogui}. The GUI follows a modular architecture, with independent threads managing concurrent acquisitions: (\textit{i}) the main thread provides visual guidance to participants; (\textit{ii}) a second thread establishes serial communication over BLE with the BioGAP NRF and retrieves raw EMG data and the US synchronization signal; (\textit{iii}) a third thread sets up TCP connection to link the MANUS C++ SDK with the Python environment, to retrieve MANUS data.
For data synchronization, the GUI appends a software-based synchronization trigger to the data streams, marking the start of acquisition and the timing of each movement: using this approach, the synchronization uncertainty can be up to $8\text{ms}$ (i.e., the sampling period of the MANUS signal), assuming ideal BLE and TCP communication. US data are streamed separately to the dedicated WULPUS GUI~\cite{frey_wulpus_2022}.

Final synchronization is performed offline using Python. US data are aligned to the EMG stream via the hardware synchronization signal. MANUS data are aligned to the EMG stream using the software trigger signal generated by the GUI. The acquisition timestamps are cross-checked to validate and confirm alignment across EMG, US, and MANUS data streams. Figure~\ref{fig:integrated_platform}(e) provides an overview of the integrated framework for the collection of EMG, US, and MANUS data.

EMG data are sampled at $500 \text{Hz}$ with a PGA gain of 6. US data are acquired from 4 channels at $8~\text{MHz}$ sampling rate with a 
pulse repetition rate of 30 Hz, resulting in $\approx133\text{ms}$ for a complete arm scan. MANUS data are acquired at $120 \text{Hz}$. 
\vspace{-0.1cm}
\subsection{Data collection protocol}
\vspace{-0.1cm}

The subject maintained a sitting position, with a $90^\circ$ elbow angle, placed on a support. Before the start of recording, a calibration procedure of the MANUS glove was executed as specified by the manufacturer.
The subject executed 11 gestures (inspired by \cite{atzori_ninapro_2015}): 6 gestures involve the activation of hand DoFs (open hand-OH, closed fist-CF, fine pinch-FP, middle pinch-MP, thumb up-TU, and thumb down-TD), 5 gestures of wrist DoFs (flexion WF, extension-WE, radial deviation-WRD, ulnar deviation-WUD, and supination-WSUP). Data are acquired in 3 sessions; between sessions, the sensors are repositioned. For US transducers, the hydrogel pads are re-applied between sessions. Each session contains 5 sets. For each set, the 11 gestures are repeated sequentially 6 times, with each position held for 4 seconds and with 1 second rest between repetitions; hence, each set consists of $\approx5\text{min}30\text{s}$ of recording. $5\text{min}$ rest is given between sessions to prevent muscle fatigue. 

\vspace{-0.1cm}
\subsection{Data Processing}
\vspace{-0.1cm}

Wrist Euler angles (flexion/extension, radial–ulnar deviation, pronation–supination) are computed from wrist quaternions provided by the MANUS Core, assuming a ZYX rotation sequence. These angles are appended to the finger joint angles (already in Euler form). Angular data are upsampled to match BioGAP's timestamps. 
A high-pass filter (Butterworth, $4^{th}$ order, $20\text{Hz}$), followed by a notch filter ($50\text{Hz}$), is applied to EMG data. No further preprocessing is performed on US.
A \textit{min-max} normalization ($[-1, 1]$ range) is applied channel-wise for both modalities. 

To construct the dataset, four US frames, corresponding to a full-arm scan from the four transducers are temporally aligned with $200~\text{ms}$ of EMG data (EMG windows slide every 4 US scans). Thus, each dataset entry comprises $200~\text{ms}$ of EMG from 8 channels (dimension $100 \times 8$), four US frames (dimension $400 \times 4$), and the corresponding 20 finger and 3 wrist joint ground-truth labels. 

\vspace{-0.1cm}
\subsection{Architectures}
\vspace{-0.1cm}

\begin{figure}[t]
    \centering
    \includegraphics[width=0.48\textwidth]{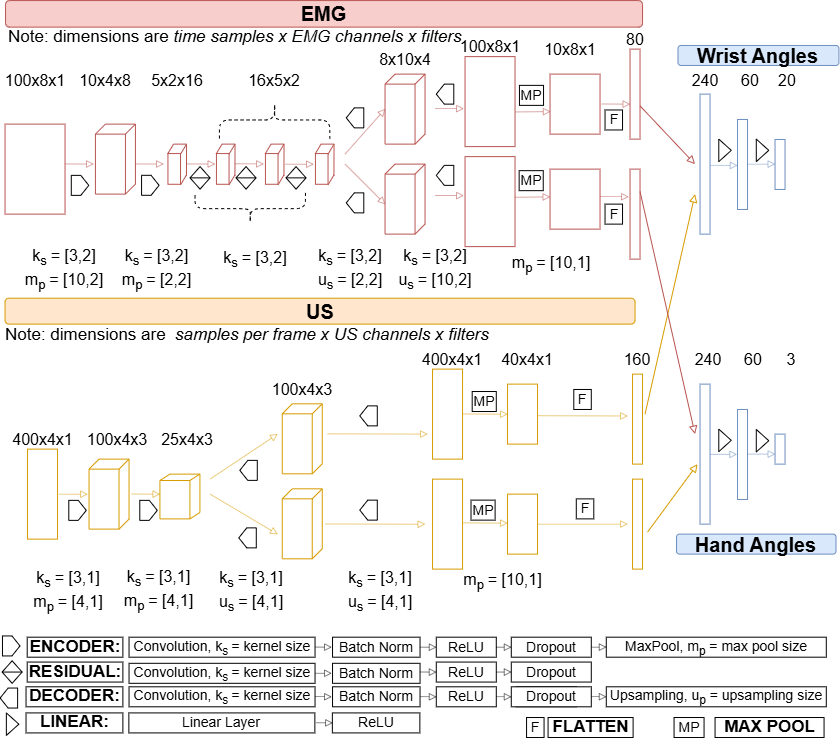}
    \vspace{-0.3cm}
    \caption{Network Architectures for EMG and US, with late-feature fusion.}
    \vspace{-0.6cm}
    \label{fig:network_architectures}
\end{figure}

\textbf{EMG architecture.} We adopt a modified version of the Neuropose architecture~\cite{liu_neuropose_2021, salter_emg2pose_2024}, with structural optimizations to reduce the footprint of the model. Also,  compared to the original Neuropose that predicted only hand joint angles, we support simultaneous prediction of both hand and wrist angles through multitask learning. We employ two encoder blocks, followed by one residual block and two decoder blocks per task. Each encoder block features a 2D convolution layer, batch normalization, ReLU activation, dropout (rate of 0.05), and max pooling. The encoder employs kernel sizes ($k_{s}$) of $[3\times2]$ and max pooling sizes ($mp_{s}$) of $[10\times 2]$ and $[2\times2]$, resulting in a temporal downsampling factor of 20 and a spatial downsampling factor of 4. This compressed representation is passed through residual blocks, consisting of three stacked convolutional layers ($k_{s}=[3\times2]$), each followed by batch normalization, ReLU, and dropout. The decoder blocks symmetrically invert the downsampling by replacing max pooling with upsampling, restoring the original temporal and spatial resolution.
While the encoder and residual layers are shared across tasks, the model includes two separate decoder branches for hand and wrist regression. After decoding, the output goes through a channel-wise max pooling operation along the time dimension, and is flattened and fed into a two-layer multilayer perceptron (MLP) to generate the final predictions for each task. The model has 14749 parameters (60KB in float32). 

\textbf{US architecture.} We use a similar encoder–decoder architecture for US, without residual connections. The encoder comprises two convolutional blocks ($k_{s}=[3\times1]$, $mp_{s}=[4\times1]$), yielding a $16\times$ spatial downsampling on a single-channel. Two task-specific decoder branches (for hand and wrist) apply symmetric upsampling ($[4\times1]$ per layer) to restore the original resolution. Final features undergo max pooling and are flattened before being passed to task-specific two-layer MLPs for joint angle regression. The model has 27773 parameters ($110\text{KB}$ in float32).

\textbf{EMG-US fusion.} We implement a late-feature fusion after the above encoder–decoder structures. First, the outputs of both modalities are flattened and concatenated. Then, this combined feature vector is processed by a task-specific MLP head (two linear layers with ReLU), which learns cross-modal interactions and outputs the final joint predictions for each task. The parameter count is 37799 (140\text{KB} in float32).  

In all analyses, the networks are trained for 25 epochs with early stopping applied if no improvement is observed on the validation set for 5 consecutive epochs. We use L2 loss function and Adam optimizer with a learning rate of $10^{-3}$ and a weight decay of $10^{-4}$. A block diagram of the model architecture integration is presented in Fig.~\ref{fig:network_architectures}. 
\vspace{-0.15cm}
\begin{figure*}[b!]
    \centering
    \vspace{-0.6cm}
    \includegraphics[width=0.90\textwidth]{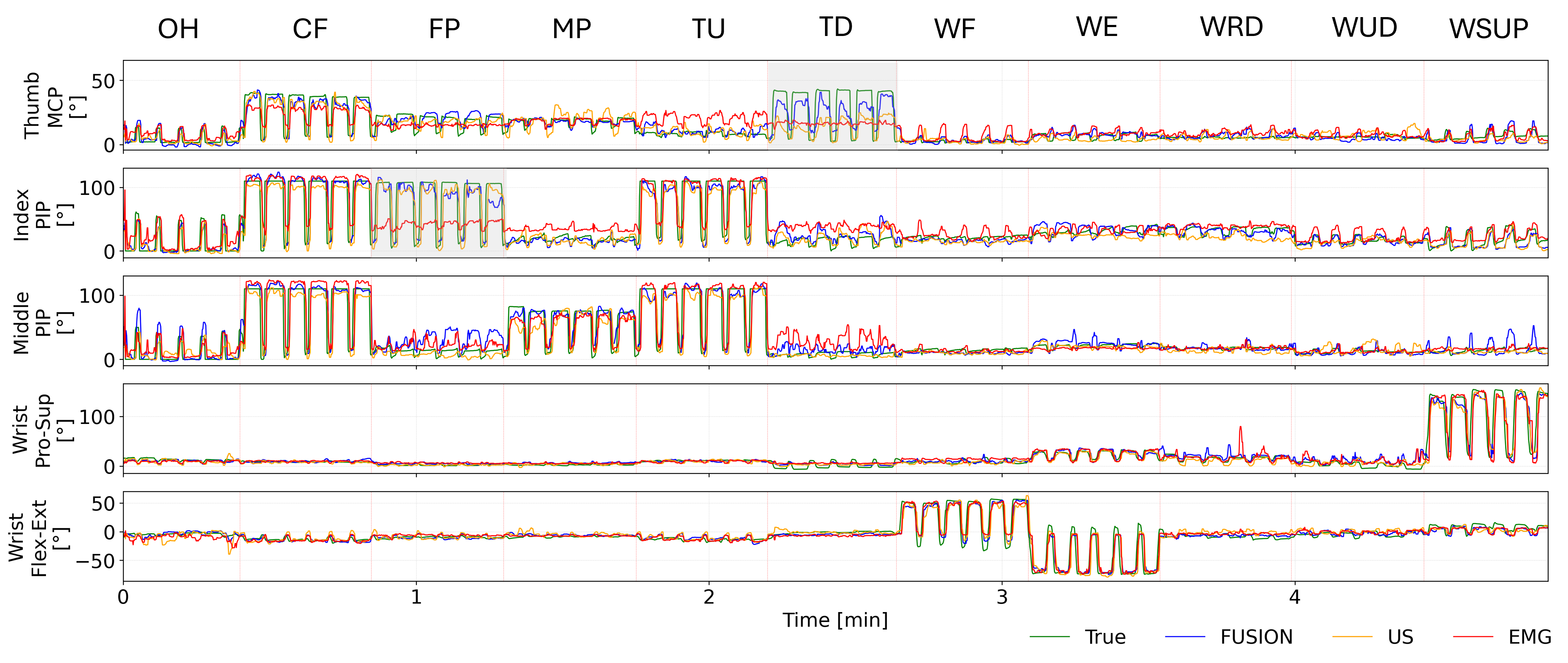}
    \vspace{-0.5cm}
    \caption{Example of ground truth vs models' predictions (smoothed with a median filter, window of $\approx670\text{ms}$) over time for one acquisition set. Light grey shades highlight the benefits of fusion when either one or both modalities fail.}
    \label{fig:predictions}
\end{figure*}

\vspace{-0.1cm}
\subsection{Evaluation procedure and performance metrics}
\vspace{-0.1cm}
In a first evaluation, which serves as a baseline, data from all three acquisition sessions are aggregated. A 5-fold cross-validation (CV) scheme is employed, where for each fold, three sets per session are used for training, one for validation, and one for testing. This setup ensures that each fold includes data from all sessions, while the test sets remain unseen during training.
In a second analysis, we adopt an intersession generalization strategy. Two sessions are used for training (with four sets per session for training and one for validation), and all sets from the remaining session are used for testing. This analysis is designed to assess model robustness under sensor repositioning. 

For both analyses, we compute the mean absolute error (MAE), root mean squared error (RMSE), and coefficient of determination (R$^2$) on each test session.

\vspace{-0.1cm}
\section{Results and Discussion}
\vspace{-0.1cm}
\textbf{System characterization. }
Our platform has dimensions of $46\text{mm}\times51\text{mm}\times15\text{mm}$ and a weight of only $14\text{g}$.
The total power consumption is $44.7\text{mW}$, with BioGAP sampling at $500\text{Hz}$ and WULPUS at $30\text{Hz}$, and both modalities streaming data via BLE. In particular, the BioGAP sub-block consumes 25.3 mW, the WULPUS sub-block consumes 17 mW, the active electrodes consume 2.4 mW, and the US-EMG hardware interfaces consumes as little as 0.05 mW.

\begin{table}[t!]
\centering
\caption{\footnotesize Error metrics} 
\vspace{-0.3cm}
\label{table:error_metrics1}

\footnotesize  
\begin{threeparttable}
\begin{tabular}{p{1.8cm}p{1.8cm}p{1.8cm}p{1.8cm}}  
\toprule[0.20em]
\textbf{Metric} & \textbf{EMG} & \textbf{US} & \textbf{EMG+US} \\

\multicolumn{4}{c}{\cellcolor{gray!20}Aggregated analyses (5-fold CV)} \\

MAE & $6.9^\circ \pm 0.5^\circ$ & $5.9^\circ \pm 0.8^\circ$ & $\textbf{5.5}^\circ \pm \textbf{0.8}^\circ$ \\
RMSE & $10.9^\circ \pm 1.0^\circ$ & $8.5^\circ \pm 1.0^\circ$ & $\textbf{8.0}^\circ \pm \textbf{1.4}^\circ$ \\
R$^2$ & $0.59 \pm 0.04$ & $0.70 \pm 0.04$ & $\textbf{0.75}\pm \textbf{0.04}$  \\
\multicolumn{4}{c}{\cellcolor{gray!20}Inter-session CV} \\
MAE & $7.8^\circ \pm 0.9^\circ$ & $9.3^\circ \pm 1.8^\circ$ & $\textbf{7.4}^\circ \pm \textbf{1.3}^\circ$ \\
RMSE & $12.0^\circ \pm 1^\circ$ & $13.1^\circ \pm 2.6^\circ$ & $\textbf{10.6}^\circ \pm \textbf{2.0}^\circ$ \\
R$^2$ & $0.54 \pm 0.03$ & $0.38 \pm 0.20$ & $\textbf{0.61} \pm \textbf{0.10}$  \\
\bottomrule[0.20em]
\end{tabular}
\vspace{-0.5cm}
\end{threeparttable}
\end{table}

\textbf{Hand Gesture Regression.}
Table~\ref{table:error_metrics1} (top rows) shows the results for the 5-fold CV. 
A per-gesture analysis (Fig.~\ref{fig:predictions}) reveals that while fusion does not consistently outperform the best single modality across all gestures, it proves advantageous in specific scenarios, such as when both modalities individually struggle (e.g., the thumb-down movement) and to compensate for failure when one modality alone performs poorly. 
\newline The intersession analysis (see Table~\ref{table:error_metrics1}, bottom rows) reveal the increased difficulty of the intersession task, which accounts for sensor displacement between sessions. 
\newline EMG-only results are comparable to SoA approaches \cite{liu_neuropose_2021},\cite{zanghieri_semg-based_2021}; US-only results are comparable with \cite{grandi_sgambato_high_2023}, \cite{spacone_tracking_2024}, but with a much higher number of DoFs.

\vspace{-0.1cm}
\section{Conclusion}
\vspace{-0.1cm}
We presented a truly-wearable ultra-low-power heterogeneous platform for concurrent EMG and US acquisition, featuring sub-$50\text{mW}$ power consumption. The platform was coupled with a novel framework for tracking full hand and wrist kinematics (23 DoFs) via EMG and US. We proposed two lightweight encoder–decoder architectures tailored for each modality with a late-feature fusion approach for their integration and a multi-task learning strategy to jointly predict 20 hand and 3 wrist DoFs. 
\review{Our system, in an intersession setting on one subject, achieved a RMSE of $10.6^\circ\pm2.0^\circ$ for the fused modalities, with $12.0^\circ\pm1^\circ$ for EMG-only and $13.1^\circ\pm2.6^\circ$ for US-only; the R$^2$ score for the fusion was $0.61\pm0.1$, with $0.54\pm0.03 $ for EMG and $0.38\pm0.20$ for US.}

\review{Future work will involve collecting data from a larger cohort of subjects, as well as conduct comprehensive statistical analyses to further validate the potential benefits of sensor fusion.}

\vspace{-0.2cm}
\section*{Acknowledgment}
We thank A. Blanco Fontao, H. Gisler, C. Leitner (ETH Z{\"u}rich) for technical support and discussions. 
\bibliographystyle{IEEEtran}
\bibliography{RelatedWorks}

\begin{thebibliography}{10}
\providecommand{\url}[1]{#1}
\csname url@samestyle\endcsname
\providecommand{\newblock}{\relax}
\providecommand{\bibinfo}[2]{#2}
\providecommand{\BIBentrySTDinterwordspacing}{\spaceskip=0pt\relax}
\providecommand{\BIBentryALTinterwordstretchfactor}{4}
\providecommand{\BIBentryALTinterwordspacing}{\spaceskip=\fontdimen2\font plus
\BIBentryALTinterwordstretchfactor\fontdimen3\font minus \fontdimen4\font\relax}
\providecommand{\BIBforeignlanguage}[2]{{%
\expandafter\ifx\csname l@#1\endcsname\relax
\typeout{** WARNING: IEEEtran.bst: No hyphenation pattern has been}%
\typeout{** loaded for the language `#1'. Using the pattern for}%
\typeout{** the default language instead.}%
\else
\language=\csname l@#1\endcsname
\fi
#2}}
\providecommand{\BIBdecl}{\relax}
\BIBdecl

\bibitem{guo_human-machine_2021}
L.~Guo, Z.~Lu, and L.~Yao, ``Human-machine interaction sensing technology based on hand gesture recognition: A review,'' \emph{{IEEE} Transactions on Human-Machine Systems}, no.~4, pp. 300--309, 2021.

\bibitem{tchantchane_review_2023}
R.~Tchantchane, H.~Zhou, S.~Zhang, and G.~Alici, ``A review of hand gesture recognition systems based on noninvasive wearable sensors,'' \emph{Advanced Intelligent Systems}, no.~10, p. 2300207, 2023.

\bibitem{shin_hgrewview_2024}
J.~Shin, A.~S.~M. Miah, M.~H. Kabir, M.~A. Rahim, and A.~Al~Shiam, ``A methodological and structural review of hand gesture recognition across diverse data modalities,'' \emph{IEEE Access}, 2024.

\bibitem{boyer_emgnoise_2023}
M.~Boyer, L.~Bouyer, J.-S. Roy, and A.~Campeau-Lecours, ``Reducing noise, artifacts and interference in single-channel emg signals: A review,'' \emph{Sensors}, no.~6, 2023.

\bibitem{yang_simultaneous_2022}
X.~Yang, Y.~Liu, Z.~Yin, P.~Wang, P.~Deng, Z.~Zhao, and H.~Liu, ``Simultaneous prediction of wrist and hand motions via wearable ultrasound sensing for natural control of hand prostheses,'' \emph{IEEE Transactions on Neural Systems and Rehabilitation Engineering}, vol.~30.

\bibitem{grandi_sgambato_high_2023}
B.~G. Sgambato \emph{et~al.}, ``High performance wearable ultrasound as a human-machine interface for wrist and hand kinematic tracking,'' \emph{IEEE Transactions on Biomedical Engineering}, 2024.

\bibitem{frey_biogap_2023}
S.~Frey, M.~Guermandi, S.~Benatti, V.~Kartsch, A.~Cossettini, and L.~Benini, ``Biogap: a 10-core fp-capable ultra-low power iot processor, with medical-grade afe and ble connectivity for wearable biosignal processing,'' in \emph{2023 IEEE International Conference on Omni-layer Intelligent Systems (COINS)}, 2023, pp. 1--7.

\bibitem{xia_toward_2019}
W.~Xia, Y.~Zhou, X.~Yang, K.~He, and H.~Liu, ``Toward portable hybrid surface electromyography/a-mode ultrasound sensing for human–machine interface,'' \emph{IEEE Sensors Journal}, vol.~19.

\bibitem{wei_multimodal_2023}
S.~Wei, Y.~Zhang, and H.~Liu, ``A multimodal multilevel converged attention network for hand gesture recognition with hybrid {sEMG} and a-mode ultrasound sensing,'' \emph{{IEEE} Transactions on Cybernetics}, vol.~53, no.~12, pp. 7723--7734.

\bibitem{yang_comparative_2020}
X.~Yang, J.~Yan, and H.~Liu, ``Comparative analysis of wearable a-mode ultrasound and {sEMG} for muscle-computer interface,'' \emph{IEEE Transactions on Biomedical Engineering}, vol.~67.

\bibitem{yin_wearable_2025}
Z.~Yin, J.~Meng, S.~Shi, W.~Guo, X.~Yang, H.~Ding, and H.~Liu, ``A wearable multisensor fusion system for neuroprosthetic hand,'' \emph{IEEE Sensors Journal}, no.~8, pp. 12\,547--12\,558.

\bibitem{krasoulis_effect_2019}
A.~Krasoulis, S.~Vijayakumar, and K.~Nazarpour, ``Effect of user practice on prosthetic finger control with an intuitive myoelectric decoder,'' \emph{Frontiers in Neuroscience}, vol.~13.

\bibitem{zanghieri_semg-based_2021}
M.~Zanghieri \emph{et~al.}, ``{sEMG}-based regression of hand kinematics with temporal convolutional networks on a low-power edge microcontroller,'' in \emph{2021 {IEEE} {COINS}}, pp. 1--6.

\bibitem{liu_neuropose_2021}
Y.~Liu, S.~Zhang, and M.~Gowda, ``{NeuroPose}: 3d hand pose tracking using {EMG} wearables,'' in \emph{Proceedings of the Web Conference 2021}.\hskip 1em plus 0.5em minus 0.4em\relax {ACM}, pp. 1471--1482.

\bibitem{salter_emg2pose_2024}
S.~Salter \emph{et~al.}, ``emg2pose: A large and diverse benchmark for surface electromyographic hand pose estimation,'' {arXiv}:2412.02725, 2025.

\bibitem{tacca_wearable_2024}
N.~Tacca, C.~Dunlap, S.~P. Donegan, J.~O. Hardin, E.~Meyers, M.~J. Darrow, S.~Colachis~Iv, A.~Gillman, and D.~A. Friedenberg, ``Wearable high-density {EMG} sleeve for complex hand gesture classification and continuous joint angle estimation,'' \emph{Scientific Reports}, no.~1, p. 18564.

\bibitem{lin_continuous_2025}
C.~Lin, C.~Zhao, J.~Zhang, C.~Chen, N.~Jiang, D.~Farina, and W.~Guo, ``Continuous estimation of hand kinematics from electromyographic signals based on power-and time-efficient transformer deep learning network,'' \emph{IEEE Transactions on Neural Systems and Rehabilitation Engineering}, vol.~33, pp. 58--67.

\bibitem{yang_wearable_2021}
X.~Yang, Y.~Zhou, and H.~Liu, ``Wearable ultrasound-based decoding of simultaneous wrist/hand kinematics,'' \emph{{IEEE} Transactions on Industrial Electronics}, vol.~68, no.~9, pp. 8667--8675, 2021.

\bibitem{spacone_tracking_2024}
G.~Spacone, S.~Vostrikov, V.~Kartsch, S.~Benatti, L.~Benini, and A.~Cossettini, ``Tracking of wrist and hand kinematics with ultra low power wearable a-mode ultrasound,'' \emph{IEEE Transactions on Biomedical Circuits and Systems}, pp. 1--13, 2024.

\bibitem{sgambato_multidofprosthetic_2024}
B.~G. Sgambato, H.~Hakami, X.~Yang, D.~Y. Barsakcioglu, A.~Jakob, M.~Fournelle, A.~H. McGregor, M.-X. Tang, and D.~Farina, ``Towards natural multi-dof prosthetic control with distributed ultrasound,'' \emph{2024 IEEE Ultrasonics, Ferroelectrics, and Frequency Control Joint Symposium (UFFC-JS)}, pp. 1--6, 2024.

\bibitem{zeng_feature_2020}
J.~Zeng, Y.~Zhou, Y.~Yang, J.~Wang, and H.~Liu, ``Feature fusion of {sEMG} and ultrasound signals in hand gesture recognition,'' in \emph{2020 {IEEE} International Conference on Systems, Man, and Cybernetics ({SMC})}, pp. 3911--3916, {ISSN}: 2577-1655.

\bibitem{pan_msmfnet_2024}
J.~Pan, J.~Chen, S.~Wei, J.~Pan, and Z.~Wang, ``{MSMFNet}: Multi-modal fusion gesture recognition network with multi-scale integration of {AUS} and {sEMG},'' in \emph{2024 International Joint Conference on Neural Networks ({IJCNN})}, pp. 1--9, {ISSN}: 2161-4407.

\bibitem{zhang_dual-modal_2025}
Y.~Zhang, S.~Wei, Z.~Wang, and H.~Liu, ``Dual-modal gesture recognition using adaptive weight hierarchical soft voting mechanism,'' \emph{IEEE Transactions on Cybernetics}, pp. 1--12.

\bibitem{frey_wulpus_2022}
S.~Frey, S.~Vostrikov, L.~Benini, and A.~Cossettini, ``Wulpus: a wearable ultra low-power ultrasound probe for multi-day monitoring of carotid artery and muscle activity,'' in \emph{2022 IEEE International Ultrasonics Symposium (IUS)}, 2022, pp. 1--4.

\bibitem{frey_ius2023}
S.~Frey, V.~Kartsch, C.~Leitner, A.~Cossettini, S.~Vostrikov, S.~Benatti, and L.~Benini, ``A wearable ultra-low-power semg-triggered ultrasound system for long-term muscle activity monitoring,'' in \emph{2023 IEEE International Ultrasonics Symposium (IUS)}, 2023, pp. 1--4.

\bibitem{ads_1298}
\BIBentryALTinterwordspacing
Ads129x low-power, 8-channel, 24-bit analog front-end for biopotential measurements. [Online]. Available: \url{https://www.ti.com/product/ADS1298\#tech-docs}
\BIBentrySTDinterwordspacing

\bibitem{AD8605_amplifier}
\BIBentryALTinterwordspacing
Precision, low noise, cmos, rail-to-rail, input/output operational amplifiers (analog devices). [Online]. Available: \url{https://www.analog.com/en/products/ad8605.html}
\BIBentrySTDinterwordspacing

\bibitem{MAX22420_digital_isolator}
\BIBentryALTinterwordspacing
Reinforced, ultra-low-power, two-channel digital isolators (analog sevices). [Online]. Available: \url{https://www.analog.com/en/products/max22420.html}
\BIBentrySTDinterwordspacing

\bibitem{datwyler_softpulse}
\BIBentryALTinterwordspacing
D.~S. Inc., ``Datwyler softpulse.'' [Online]. Available: \url{https://datwyler.com/company/innovation/softpulse/products}
\BIBentrySTDinterwordspacing

\bibitem{hydrogel_pads}
\BIBentryALTinterwordspacing
Ems gel pad. Accessed via Amazon ADAKEL EMS Gel Pad product page. [Online]. Available: \url{https://amzn.eu/d/9dQ3nRG}
\BIBentrySTDinterwordspacing

\bibitem{manus_quantum_glove}
\BIBentryALTinterwordspacing
M.~T. Group. [Online]. Available: \url{https://www.manus-meta.com/products/quantum-metagloves}
\BIBentrySTDinterwordspacing

\bibitem{saggio_quasi-static_2025}
\BIBentryALTinterwordspacing
G.~Saggio, L.~Pietrosanti, I.-J. Lee, and B.-S. Lin, ``Quasi-static and dynamic measurement performances of an electromagnetic field-based sensory glove termed manus quantum metaglove.'' [Online]. Available: \url{https://dx.doi.org/10.2139/ssrn.5208853}
\BIBentrySTDinterwordspacing

\bibitem{biogui}
M.~Orlandi, P.~M. Rapa, M.~Zanghieri, S.~Frey, V.~Kartsch, L.~Benini, and S.~Benatti, ``Real-time motor unit tracking from semg signals with adaptive ica on a parallel ultra-low power processor,'' \emph{IEEE Transactions on Biomedical Circuits and Systems}, no.~4, pp. 771--782, 2024.

\bibitem{atzori_ninapro_2015}
M.~Atzori, A.~Gijsberts, I.~Kuzborskij, S.~Elsig, A.-G. Mittaz~Hager, O.~Deriaz, C.~Castellini, H.~Müller, and B.~Caputo, ``Characterization of a benchmark database for myoelectric movement classification,'' \emph{IEEE Transactions on Neural Systems and Rehabilitation Engineering}, vol.~23, no.~1, pp. 73--83, 2015.

\end{thebibliography}
\end{document}